\documentclass[letterpaper,twocolumn,10pt]{article}

\usepackage{usenix,epsfig,endnotes,multirow}
\usepackage{listings}
\usepackage{color}
\usepackage{amsmath} 
\usepackage{bm}
\usepackage{citesort}
\usepackage{graphicx}
\usepackage{url}

\usepackage{caption}
\usepackage{titlesec}
\usepackage[caption=false]{subfig}

\definecolor{myred}{rgb}{0.9,0.1,0.1}
\definecolor{mygray}{rgb}{0.5,0.5,0.5}
\definecolor{mymauve}{rgb}{0.58,0,0.82}

\newcommand{\nameS}[0]{{\em BinderCracker} }
\newcommand{\name}[0]{{\em BinderCracker}}
\newcommand{\binder}[0]{{\tt Binder}}

\begin{document}
		

\title{\Large \bf BinderCracker: Assessing the Robustness of Android System Services}

\author{
{\rm Huan Feng, Kang G. Shin}\\
Department of Electrical Engineering and Computer Science\\
The University of Michigan\\
\{huanfeng, kgshin\}@umich.edu
} 

\maketitle


\begin{abstract}

In Android, communications between apps and system services are supported by a 
transaction-based Inter-Process Communication (IPC) mechanism. \binder, as 
the cornerstone of this IPC mechanism, separates two communicating parties as 
client and server. As with any client--server model, the server should not make 
any assumption on the validity (sanity) of client-side transaction. To our 
surprise, we find this principle has frequently been overlooked in the implementation 
of Android system services. In this paper, we demonstrate the prevalence and severity 
of this vulnerability surface and try to answer why developers keep making this 
seemingly simple mistake. Specifically, we design and implement \name, 
an automatic testing framework that supports parameter-aware fuzzing and has 
identified more than 100 vulnerabilities in six major versions of Android, including 
the latest version Android 6.0, Marshmallow. Some of the vulnerabilities have severe 
security implications, causing privileged code execution or permanent Denial-of-Service 
(DoS). We analyzed the root causes of these vulnerabilities to find that most of them 
exist because system service developers only considered exploitations via public APIs. 
We thus highlight the deficiency of testing only on client-side public APIs and argue 
for the necessity of testing and protection on the \binder\ interface --- the actual 
security boundary. Specifically, we discuss the effectiveness and practicality of 
potential countermeasures, such as precautionary testing and runtime diagnostic.

\end{abstract}


\section{Introduction}

Android is the most popular smartphone OS and dominates the global market 
with a share of more than 82\%~\cite{AndroidShares}. By the end of 2015, the 
total number of Android devices surpassed 1.4 billion, and there were more than 
1.6 million mobile apps available in Google Play for download~\cite{AndroidAppNumbers,AndroidDeviceNumbers}. 
The developers of these apps are not always trustworthy; many of them might 
be inexperienced, careless or even malicious. Therefore, proper isolation 
between apps and the system is essential for robustness and security.

To meet this requirement, apps in Android execute in application sandboxes. If 
an app wants to interact with system services, it must perform Inter-Process 
Communications (IPC). \binder, as the cornerstone of this IPC mechanism, separates two 
communicating parties as client and server. Each server (i.e., system service in this case) 
exports a list of public APIs that clients (i.e., mobile apps) can invoke. The 
input parameters of each API go through extensive sanity checks on the client-side 
before being packed into a transaction and then sent to the server. An AIDL 
(Android Interface Description Language) file further enforces the schema of each 
transaction, serving as an explicit contract between the client and the server. 
These client-side public APIs get examined and tested by thousands of participating 
vendors of AOSP (Android Open Source Project) and are believed to be robust.

Robust public APIs and AIDL enforcement can protect system services against erroneous 
input from careless or inexperienced app developers, but render ineffective against 
developers with an adversarial mindset. This is because both enforcements reside in 
the process space of the client (attacker) and thus can eventually be circumvented. 
In other words, any system service that hinges on the validity (sanity) of client-side 
transaction is fundamentally vulnerable --- the server should be robust 
on its own. This is probably a best engineering practice for any system that adopts 
a client--server model, but has surprisingly been overlooked in the implementation of 
many Android system services. In this paper, we conduct a comprehensive study of 
this attack surface in Android. Specifically, we answer two important questions: 
\begin{description}
\item[Q1.] {\it How prevalent is this issue in Android, a major open source project?}
\item[Q2.] {\it Why are so many developers making this seemingly simple mistake?}
\end{description}

To answer these questions, we designed and implemented \name, an automatic 
testing framework for Android system services. \nameS automatically crawls the 
RPC interfaces of both Java and native system services, and injects fuzzing 
transactions via the \binder\ surface. This directly challenges the robustness 
of the error-handling mechanisms in the system services. To increase the scale 
and depth of our testing, \nameS supports parameter-aware fuzzing with semi-valid 
inputs. This requires us to record and mutate existing transactions and understand 
their semantics. Doing this in the context of Android is challenging because 
a transaction may contain remote object handles that cannot be recorded in the form of 
raw bytes. Moreover, transactions frequently contain dynamic and non-primitive data 
types that are difficult to mutate in a sensible way. \nameS overcomes these challenges 
by implementing a replay engine that has in-depth understanding of \binder\ transactions. 
It utilizes the dependencies between transactions to reconstruct remote object handles 
during runtime, and tracks the hierarchy of non-primitive data types to unmarshall them 
into primitive types.

We examined more than 2400 service APIs in 6 major versions of Android, including the 
latest Android 6.0 (Marshmallow). In total, we identified more than 100 vulnerabilities, 
most of which are unfixed to date. Many of the vulnerabilities we identified are found 
to be able to crash the entire Android Runtime, while others can cause specific system 
services or system apps to fail. Some vulnerabilities have severe security implications, 
and may result in system memory corruption, privileged code execution, targeted or 
permanent Denial-of-Service (DoS). This extensive testing also demonstrates the effectiveness 
of our parameter-aware fuzzing --- it identified 7x more vulnerabilities than simple 
black-box fuzzing with the same amount of time. Furthermore, since \nameS is parameter-aware, 
we can now unearth different vulnerabilities in the same RPC method by fuzzing different 
parameters in the same transaction.

We further analyzed the root causes of the identified vulnerabilities and studied 117 
of them in Android source codes. Most of them exist because the developers only 
considered exploitations of public APIs. Therefore, many risky scenarios are assumed 
to be `unlikely' or even `impossible' in their mindset. Here, we list three most common 
mistakes made by system service developers that contribute to a vast majority of the 
vulnerabilities. First, private APIs are assumed to be unknown to others, thus no 
sanity-check is made. Second, client-side enforcements are assumed to be secure, so 
there is no double check of them at the server-side. Third, the de-serialization process 
is assumed to be always undisturbed, hence no sanity check is conducted during this process. 
All of these are unreliable assumptions because they either directly or indirectly depend 
on the validity of client-side transactions. Moreover, we demonstrate most of the 
vulnerabilities can never be found by testing only public APIs. We, therefore, highlight 
the deficiency of testing only client-side public APIs and argue for the necessity of 
testing and protection at the Binder surface --- the actual security boundary.

As our findings indicate, new vulnerabilities keep on emerging, especially with releases 
of new Android versions. To address this emerging attack surface, we need to eliminate 
potential vulnerabilities as early as possible in the development cycle. Specifically, we suggest 
the use of various precautionary testing techniques (including \name) before each product 
release. This can stop a large number of vulnerabilities from reaching the end-users. 
In fact, many severe vulnerabilities~\cite{peles2015one,cve1528,cve1474} could have been avoided, 
had \nameS been deployed. We also describe how to enhance the visibility of \binder\ 
transaction during runtime to support more informative runtime diagnosis, in case 
some vulnerabilities leak through \nameS and eventually reach the end-users.

This paper makes five main contributions by:

\begin{itemize}

\item Conducting the first comprehensive study that assesses the robustness 
of Android system services and unveiling an alarming perspective.

\item Designing and implementing \name, an automatic testing framework 
that supports parameter-aware fuzzing on the \binder\ surface; 

\item Conducting an extensive test on 6 major Android versions and 
identifying 100+ vulnerabilities, some of which have severe security implications; 

\item Unearthing the root causes of these vulnerabilities and highlighting 
the necessity of protection on \binder\ --- the actual security boundary;

\item Discussing the effectiveness and practicality of potential countermeasures, 
such as precautionary testing and runtime diagnostic techniques.

\end{itemize}

The rest of the paper is organized as follows. Section 2 summarizes related work 
in the field of software testing and Android security. Section 3 introduces 
\binder\ and AIDL in Android, and describes how Android uses these to build system services. 
Section 4 examines the attack surface under our investigation. Section 5 details the design 
and implementation of an automatic testing framework, \name, which exposes vulnerable system 
services. Section 6 describes our tests on stock Android firmware and analyzes the root 
causes and security implications of the discovered vulnerabilities. Section 7 gives a 
comprehensive discussion on how to effectively eliminate these vulnerabilities in the 
development cycle. Section 8 discusses the long-term value and other potential 
uses of our work, and finally, the paper concludes with Section 9.


\section{Related Work} 

Discussed below is related work in the field of software testing and 
Android security.

\paragraph{Software Testing.} In the software community, robustness 
testing falls into two categories: {\em functional} and {\em exceptional} 
testing. Functional testing focuses on verifying the functionality of 
software using expected input, while exceptional testing tries to apply 
unexpected and faulty inputs to crash the system. Numerous efforts have been 
made in the software testing community to test the robustness of 
Android~\cite{Ye:2013:DFA:2536853.2536881,amalfitano2012using,hu2011automating,%
mcdonnell2013empirical,amalfitano2011gui,machiry2013dynodroid}. Most of them 
focus on the functional testing of GUI elements~\cite{amalfitano2012using,hu2011automating,amalfitano2011gui,machiry2013dynodroid}. 
Some have conducted exceptional testing on the evolving public APIs~\cite{mcdonnell2013empirical}. 
In this paper, we highlight the deficiency of testing only on public APIs and 
conduct an exceptional testing on lower-level \binder-based RPC interfaces.

\paragraph{Android Security.} Android has received significant attention from the research community as 
an open source operating system~\cite{Enck:2010:TIT:1924943.1924971,%
Rangwala:2014:TPE:2582557.2582562,bugiel2011xmandroid,marforio2011application,%
Hornyack:2011:TAD:2046707.2046780,Enck:2011:SAA:2028067.2028088,Shabtai:2010:GAC:1803940.1804131}. 
Existing Android security studies largely focus on the imperfection of 
high-level permission model~\cite{Felt:2011:APD:2046707.2046779,Felt:2012:APU:2335356.2335360,%
Nauman:2010:AEA:1755688.1755732}, and the resulting issues, such as information leakage~\cite{Enck:2010:TIT:1924943.1924971}, privilege escalation~\cite{Rangwala:2014:TPE:2582557.2582562,bugiel2011xmandroid} 
and collusion~\cite{marforio2011application}. Our work highlights the 
insufficient protection of Android's lower-level \binder-based RPC 
mechanism and how it affects the robustness of system services.

There also exist a few studies focusing on the IPC mechanism of Android
\cite{Maji:2012:ESR:2354410.2355135,Chin:2011:AIC:1999995.2000018,Sasnauskas:2014:IFC:2632168.2632169,%
Maji:2012:ESR:2354410.2355135,Kantola:2012:RAS:2381934.2381948,elish2015need}. 
However, they largely focus on one specific instance of Android IPC --- Intent. Since 
the senders and recipients of Intents are both apps, manipulating Intents will not serve 
the purpose of exposing vulnerabilities in system services. Some researchers also provide 
recommendations for hardening Android IPCs~\cite{Kantola:2012:RAS:2381934.2381948,%
Maji:2012:ESR:2354410.2355135} and point out that the key issue in Intent communication 
is the lack of formal schema. We demonstrate that even for mechanisms enforcing a formal 
schema, such as AIDL, robustness remains as a critical issue. Gong {\em et al.}~\cite{GuangGong} 
also conducted experiments on the fuzzing of \binder\ interface. However, they focused on 
implementing Proof-of-Concept (PoC) exploits using identified vulnerabilities, instead of 
comprehensively assessing and understanding the attack surface. In fact, they only tested 
simple black-box fuzzing on one Android version, which is a small subset of our work. We 
regard their work as a parallel and independent effort from an industry perspective.


\section{Android IPC and Binder}

Android executes apps and system services as different processes and 
enforces isolation between them. To enable different processes to 
exchange information with each other, Android provides, \binder, a 
secure and extensible IPC mechanism. Described below are the basic 
concepts in the \binder\ framework and an explanation of how a 
typical system service is built using these basic primitives.

\subsection{Binder}

In Android, \binder\ provides a message-based communication channel between two processes. 
It consists of (i) a kernel-level driver that achieves communication 
across process boundaries, (ii) a \binder\ library that uses \texttt{ioctl} 
syscall to talk with the kernel-level driver, and (iii) upper-level abstracts 
that utilize the \binder\ library. Conceptually, \binder\ takes a classical 
client--server architecture. A client can send a transaction to the remote server via the 
\binder\ framework and then retrieves its response. The parameters of the transaction are 
marshalled into a \texttt{Parcel} object which is a serializable data container. 
The \texttt{Parcel} object is sent through the \binder\ driver and then gets delivered to the 
server. The server de-serializes the parameters of the \texttt{Parcel} object, processes 
the transaction, and returns a response in a similar way back to the client. This allows 
a client to achieve {\it Remote Procedure Call} (RPC) and invoke methods on remote 
servers as if they were local. This \binder-based RPC is one of the most frequent 
forms of IPC in Android, and underpins the implementation of most system services.

\subsection{AIDL}

Many RPC systems use IDL (\textit{Interface Description Language}) to define 
and restrict the format of a remote invocation~\cite{Maji:2012:ESR:2354410.2355135}, 
and so does Android. The AIDL (\textit{Android Interface Description Language}) 
file allows the developer to define the RPC interface both the client and the server 
agree upon~\cite{AndroidAIDL}. Android can automatically generate Stub and Proxy 
classes from an AIDL file and relieve the developers from (re-)implementing the 
low-level details to cope with native \binder\ libraries. The auto-generated Stub and Proxy 
classes will ensure that the declared list of parameters will be properly 
serialized, sent, received, and de-serialized. The developer only needs to provide 
a \texttt{.aidl} file and implement the corresponding interface. In other words, 
the AIDL file serves as an explicit contract between client and server. This enforcement 
makes the \binder\ framework extensible, usable, and robust. Fig.~\ref{aidl} shows an 
example AIDL file that defines the interface of a service that implements a queue.

\begin{figure}[t]

\lstset{language=Java}

\begin{lstlisting}
interface IQueueService {
	boolean add(String name);
	String peek();
	String poll();
	String remove();
}
\end{lstlisting}
\caption{An example AIDL file which defines the interface of a service 
that implements a queue.}
\label{aidl}
\end{figure}

\subsection{System Service}

\begin{figure}
	\centering
	\hspace{5mm}
	\includegraphics[scale = 0.2]{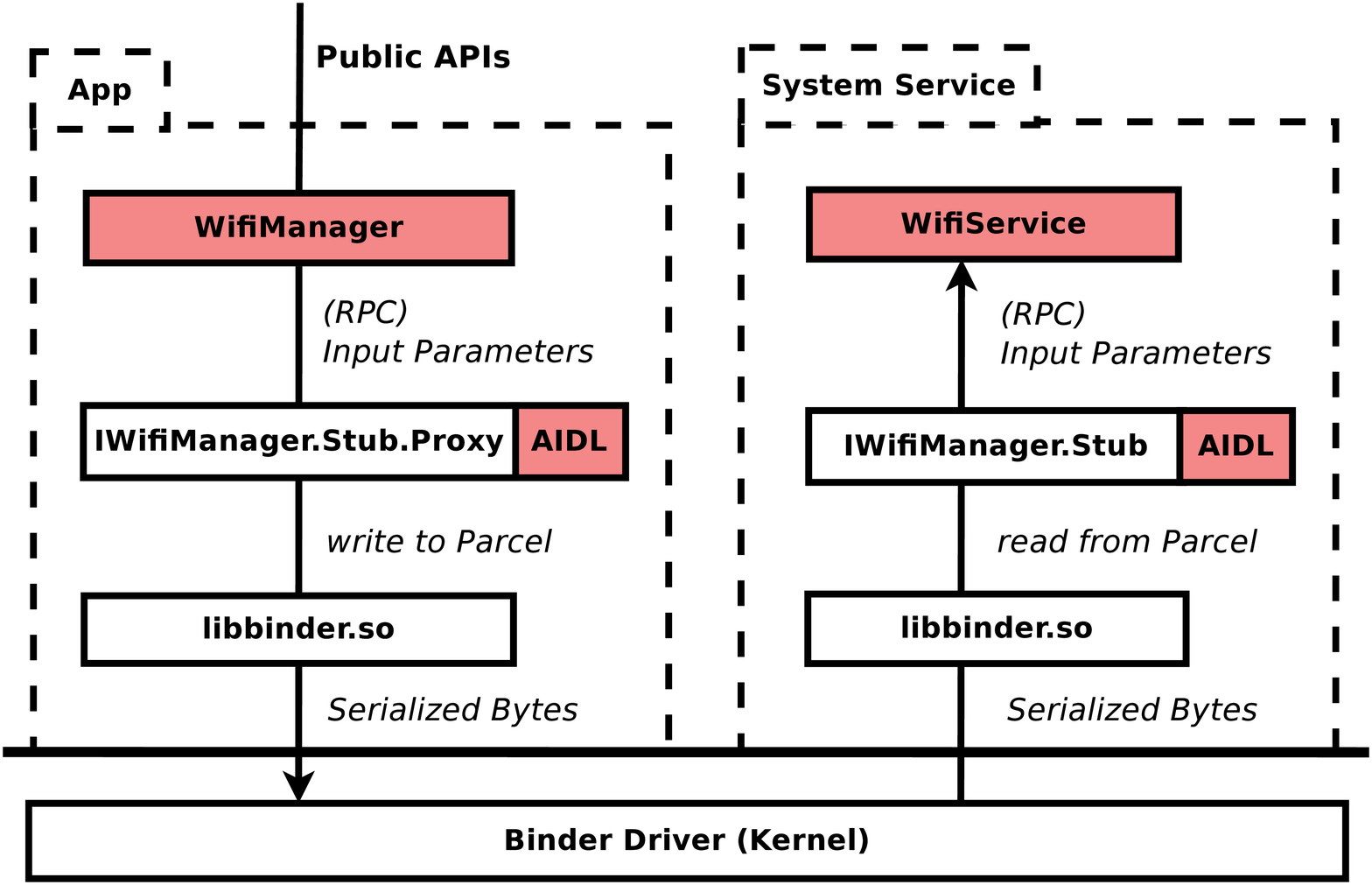}
	\caption{\small How does an app communicate with a system service using 
	\binder-based RPC (using Wi-Fi service as an example)? 
	The red shaded region represents the codes that need to be 
	provided/implemented by the service developer. }
	\label{service}
\end{figure}

We now describe how the low-level concepts in the \binder\ framework are structured 
to deliver a system service, using Wi-Fi service as an example. To implement 
the Wi-Fi service, system developers only need to define its interfaces as an 
AIDL description, and then implement the corresponding server-side 
logic (\texttt{WifiService}) and client-side wrapper (\texttt{WifiManager}) 
(see Fig.~\ref{service}). The serialization, transmission, and de-serialization 
of the interface parameters are handled by the codes automatically generated from 
the AIDL file. Specifically, when the client invokes some RPC method in the client-side 
wrapper \texttt{WifiManager}, the Proxy class \texttt{IWifiManager.Stub.Proxy} will 
marshall the input parameters in a \texttt{Parcel} object and send it across the process 
boundary via the \binder\ driver. The \binder\ library at the server-side will then 
unmarshall the parameters and invoke the \texttt{onTransact} function in the Stub 
class \texttt{IWifiManager.Stub}. This eventually invokes the service logic programmed 
in \texttt{WifiService}. Fig.~\ref{service} provides a clear illustration of the entire process.

\section{The Attack Surface}

The \binder\ framework separates two communicating parties as client and server. In 
the most common scenario, the client is a user-level app and the server is a system 
service. As with any client--server model, the server should never trust the client. 
In the particular case of Android which uses the \binder\ framework to build a 
light-weight RPC mechanism, the following two important properties have to be guaranteed.

\begin{itemize}

\item The RPC interfaces in both client and server sides should be consistent: 
they should expect the same list of input and return parameters.

\item Each parameter of the RPC interface should be properly checked: the server 
should only accept a transaction if all of its parameters are valid.

\end{itemize}

To guarantee these properties, Android adopts an AIDL enforcement and conducts extensive 
testings on public APIs. AIDL serves as an explicit contract between the server and the 
client. It defines the RPC interfaces a service trying to provide. A system service developer can work 
on top of an AIDL interface and leave the serialization, transmission and de-serialization 
to be handled automatically by the codes generated from the AIDL file. Bugs in the codes that are 
manually written by the developer are eliminated further by testing on public APIs and feedbacks 
from thousands of vendors and hundreds of millions of users. Thanks to these mechanisms, Android 
services, especially the widely-used and well-maintained system services, could be made robust.

\begin{figure}
	\centering
	\hspace{5mm}
	\includegraphics[scale = 0.2]{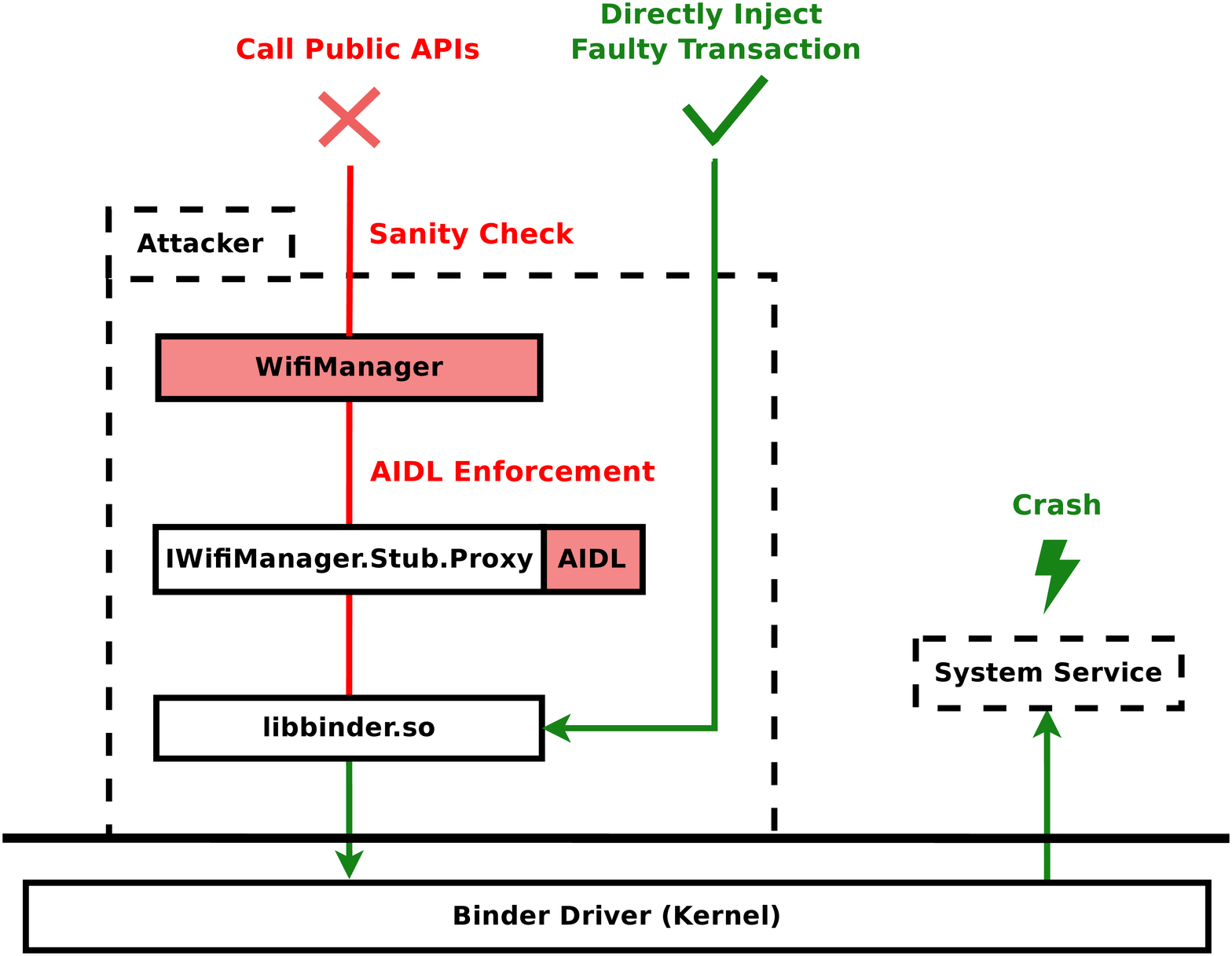}
	\caption{\small By injecting faulty transactions via the \binder\ driver, 
	an attacker can bypass the protections in the upper layers of the program stack, 
	and directly confront the server-side exception handling mechanisms.}
	\label{attack}
\end{figure}

However, when the client is malicious and trying to sabotage system services intentionally, 
these mechanisms cannot be counted on as a secure measure. This is because both enforcements 
reside in the process space of the client (attacker), thus can eventually be circumvented. 
Specifically, although AIDL enforcement and public API testing consolidate the upper and middle 
layers of the program stack, the service APIs (RPC interfaces) are still directly exposed in the 
low-layer of the stack --- the \binder\ driver. By directly injecting faulty transactions into 
the \binder\ driver, an attacker can circumvent these existing protections in the upper layers 
and directly confront the server-side exception handling mechanisms. Fig.~\ref{attack} gives 
an illustrative view of this attack surface.

Ideally, this should not affect the robustness of system services. The server-side codes should 
be robust on its own without making any assumption on the client-side transactions. This is probably a 
best engineering practice for any system that adopts a client--server model. However, we found 
this principle frequently overlooked in the implementation of Android system services, thus motivating 
us to perform a comprehensive study of this attack surface. Specifically, 
we want to answer two questions: {\it (1) how prevalent is this issue in Android, a major open source 
project? and (2) why are so many developers making this seemingly simple mistake?}

\subsection{Attack Model}

In this paper, we assume the adversary is a malicious app developer trying to sabotage system 
services. The adversary may mount this attack for various malicious purposes, such as 
launching a Denial-of-Service (DoS) attack, achieving privileged code execution, etc. A system 
service can be generic, existing in Android framework base, or vendor-specific, introduced by 
device manufacturers. The attacker has no root permission and cannot penetrate the security 
of OS kernel. 


\section{Automatic Vulnerability Discovery}

To answer the questions raised above, we design and implement an automatic 
testing framework, \name, that can effectively unearth vulnerabilities in system 
services. To the best of our knowledge, \nameS is the first tool that 
supports parameter-aware fuzzing on the \binder\ surface.

\subsection{BinderCracker: An Overview}

To fuzz a system service, \nameS must be able to send mal-formatted 
transactions to it. The attack surface we use is fundamental --- the 
client can send any arbitrary transaction to the server because the client 
has complete control over its own process space, and can thus bypass any 
client-side enforcement. This can be achieved by either taking 
advantage of hidden Android APIs or hijacking the libc call that 
underpins the \binder\ communication library. Both of these techniques 
can be achieved in user-level without extending the Android 
system~\cite{Xu:2012:APP:2362793.2362820}. Basically, 
\nameS is manipulating (either directly or indirectly) a 
\texttt{binder\_transaction\_data} struct sent to the \binder\ driver. 
This data struct contains three important pieces of information we need 
to modify to send a fuzzing transaction and has the format as shown in 
Fig.~\ref{btd}.

\begin{figure}[t]
\lstset{language=C++}
\begin{lstlisting}
struct binder_transaction_data {
	union {
		size_t handle; // (1).target service
		void *ptr;
	}target;
	void *cookie;
	unsigned int code; // (2).RPC method
	unsigned int flags;
	pid_t sender_pid;
	uid_t sender_euid;
	size_t data_size; 
	size_t offsets_size; 
	union {
        struct {
        	 
           binder_uintptr_t buffer;
           binder_uintptr_t offsets;
        } ptr;
        __u8 buf[8];
    } data; // (3).transactional data
};
\end{lstlisting}
\caption{\small The data struct sent through the \binder\ diver via the \texttt{ioctl} 
libc call. This struct contains three important pieces of information we need to modify to 
send a fuzzing transaction.}
\label{btd}
\end{figure}

The \texttt{target.handle} field specifies the service this transaction 
is sent to. The \texttt{code} field represents a specific RPC method we want 
to fuzz. The {\texttt data} struct contains the serialized 
bytes of the list of parameters for the RPC method, which is inherently 
a \texttt{Parcel} object. \texttt{Parcel} is a container class that provides 
a convenient set of serialization and de-serialization methods for different 
data types. Both the client and the server work directly with this \texttt{Parcel} 
object to send and receive the input parameters. Later in this section, 
we will elaborate on how to modify the \texttt{handle} and \texttt{code} variables to redirect 
the transaction to a specific RPC method of a specified service, and how to fuzz 
the \texttt{Parcel} object to facilitate testing with different policies.

\subsection{Transaction Redirection}

There is a one-to-one mapping from the \texttt{handle} variable in 
the \texttt{binder\_transaction\_data} object to system service. 
This mapping is created during runtime and maintained by the \binder\ driver. 
Since the client has no control over the \binder\ driver, it cannot get 
this mapping directly. For system services that are statically cached, we can get 
them indirectly by querying a static service manager which has a fixed \texttt{handle} 
of 0. This service manager is a centralized controller for service registry 
and will be started before any other services. By sending a service interface 
descriptor (such as android.os.IWindowManager) to the service manager, 
it will return an \texttt{IBinder} object which contains the \texttt{handle} 
for the specified service. For system services that are dynamically 
allocated, we can retrieve them by recursively replaying the supporting 
transactions that generate these services. We will elaborate this later 
when discussing transaction fuzzing.

After getting the \texttt{handle} of a system service, we need to further specify the 
\texttt{code} variable in the \texttt{binder\_transaction\_data} object. 
Each code represents a different RPC method defined in the AIDL file. 
This mapping can be found in the Stub files which are automatically 
generated from the AIDL file. The \texttt{code} variable typically ranges from 1 
to the total number of methods declared in the AIDL file. For native 
system services that are not implemented in Java, this mapping is 
directly coded in either the source files or the header files. 
Therefore, we scan both the AIDL files and the native source codes of 
Android to construct the mapping between transaction codes and RPC methods.

\subsection{Transaction Fuzzing}

After being able to redirect a \binder\ transaction to a chosen RPC method of a 
chosen system service, the next step is to manipulate the transaction data and 
create faulty transactions that are unlikely to occur in normal circumstances. 
Here, we take three widely-used fuzzing policies: sending empty transaction, 
random transactions, and semi-valid transactions. The first two policies are easy 
to implement because it is agnostic of the RPC method we target --- we only need 
to fill the transaction with either NULL values or randomly generated bytes. 
The last policy, however, requires us to understand the semantics of a transaction 
to fuzz each parameter individually. Next, we explain how \nameS supports parameter-aware 
fuzzing with semi-valid transactions and why it is challenging even when we already 
know the RPC interfaces.

\paragraph{Parameter-Aware Fuzzing.} To increase the scale and depth of testing, 
\nameS supports fuzzing with semi-valid transactions. A transaction is said to 
be {\em semi-valid} if all of the parameters it contains are valid except for one. 
Semi-valid transactions can dive deeper into the program structure without being 
early rejected, thus is able to reveal more in-depth vulnerabilities. To test 
with semi-valid transactions, we need to first record valid (seed) transactions, 
and then mutate the parameters in each transaction. This requires \nameS to be 
parameter-aware and is challenging for two reasons. First, recording a transaction 
is challenging when the transaction involves remote objects that cannot be recorded 
as raw bytes. In this scenario, values in the raw bytes are merely handles to the 
remote objects and become meaningless once out of the current execution context. 
Second, mutating a transaction is challenging when the transaction contains dynamic 
or non-primitive data types. Since the internal structure of this data type is unknown, 
we do not know how to mutate it in a sensible way. For example, many RPC interfaces take 
\texttt{Intent} as an input parameter. As a non-primitive data type, an \texttt{Intent} may 
contain arbitrary types of primitive types (i.e., \texttt{Int}, \texttt{String}, \texttt{Double}), 
depending on what has been put into it during runtime. Next, we will detail how 
we overcome these technical challenges.

\begin{figure}
	\centering
	\hspace{5mm}
	\includegraphics[scale = 0.2]{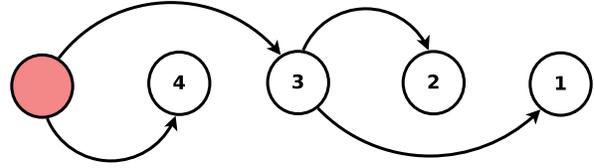}
	\caption{\small When fuzzing a transaction, we need to replay the supporting 
	transactions according to their relative order in the dependency graph. This way, 
	all the remote objects this transaction requires will be reconstructed during runtime. }
	\label{dependency}
\end{figure}

\begin{figure*}
	\centering
	\hspace{5mm}
	\includegraphics[scale = 0.2]{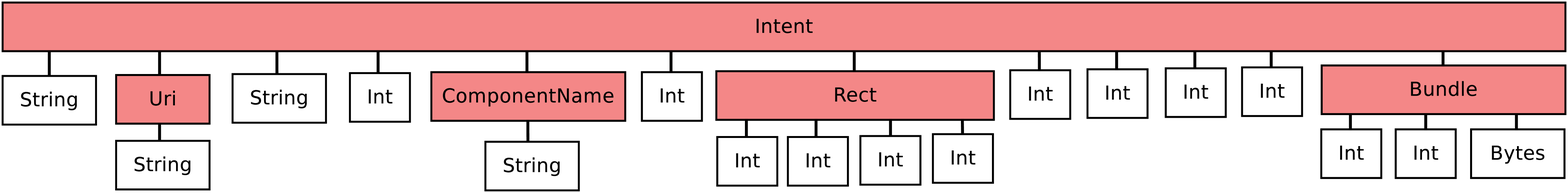}
	\caption{\small The internal type structure of a non-primitive data type, 
	\texttt{Intent}, generated by recording the de-serialization process of 
	each non-primitive type. Note that this type structure is dynamic --- it 
	depends on what has been put into this \texttt{Intent} during runtime.}
	\label{intent}
\end{figure*}

\paragraph{Remote Object.} In Android, more than 14\% of RPC methods and 37\% of 
user-level RPC calls involve a remote object. The most frequent form of a 
remote object is an \texttt{IBinder} object, which is widely-used for registering and 
invoking remote callbacks. Recording the raw bytes of these objects won't work since 
they are merely object handles. We overcome this challenge by maintaining a dependency 
graph among transactions. When recording each transaction, we iterate through the 
list of remote objects it takes as input and generates as output. Then, we construct 
a dependency graph that records how dynamic \texttt{IBinder}s are produced and consumed. 
Before trying to replay a transaction, we need to execute the supporting transactions 
according to their relative order in the dependency graph (see Fig. 5). This way, all the 
remote objects this transaction requires will be reconstructed and cached beforehand. A 
similar technique is also used to generate the handle of dynamically generated system services.

\paragraph{Non-primitive Data Types.} In Android, more than 48\% of the RPC methods 
involve non-primitive data types. Since we do not know their internal type structures, 
we cannot effectively fuzz it. We solve this problem by instrumenting the (de-)serialization 
functions in the \texttt{Parcel} class. During the recording process of the seed transaction, 
when the client de-serializes each input parameter from the \texttt{Parcel} object 
(the transaction), we also record its hierarchical meta-data by recording the orders 
of the function invocations. This way, we know how to unmarshall every non-primitive 
data types and can decompose a seed transaction into an array of primitive types. We 
then iterate through this list and mutate each unmarshalled primitive types. For 
numerical types such as Integer, we may add or substrate a small delta from the current 
value or change it to Integer.MAX, 0 or Integer.MIN; for literal types such as String, 
we may randomly mutate the bytes contained in the String or insert special \
characters at certain locations. Fig.~\ref{intent} illustrates the internal type 
structure of a non-primitive data type, \texttt{Intent}, generated by recording 
its de-serialization process.

\begin{figure}
	\centering
	\hspace{5mm}
	\includegraphics[scale = 0.2]{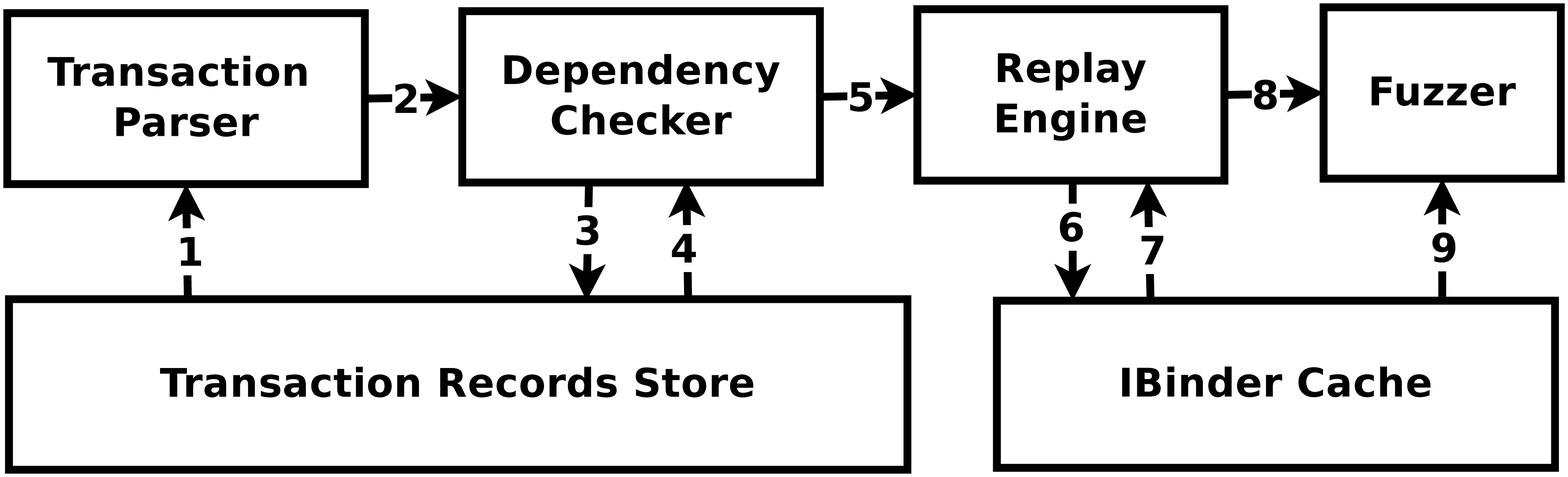}
	\caption{\small How does \nameS generate semi-valid fuzzing transactions from seed transactions.}
	\label{bindercracker}
\end{figure}

In summary, \nameS maintains both the type hierarchy and dependency graph 
when recording a seed transaction. These information capture the semantic 
and context of each transaction and help \nameS generate semi-valid fuzzing 
transactions. Specifically, it follows the process illustrated in 
Fig.~\ref{bindercracker}. For each seed transaction we want to fuzz, we first 
parse the raw bytes of the transaction and unmarshall non-primitive data types 
into an array of primitive types (step 1). This utilizes the type hierarchy recorded 
with the seed transaction. Then, we check the dependency of the transaction 
(step 2) and retrieve all the supporting transactions (steps 3, 4). This step 
utilizes the dependency graph recorded with the seed transaction. After that, 
we need to replay the supporting transactions (step 5) to generate and cache 
the remote \texttt{IBinder} object handles (steps 6, 7). Finally, the fuzzer 
can start to generate semi-valid fuzzing transactions by mutating each parameter 
in the seed transaction (steps 8, 9).

\subsection{Server Exceptions}

After sending a faulty transaction to a remote service, there 
are a few possible responses from the server-side. First, the server 
detects the input is invalid and rejects the transaction, writing an 
\texttt{IllegalArgumentException} message back to the client. Second, the server 
accepts the argument and starts the transaction, but encounters 
unexpected states or behaviors and catches some type of \texttt{RuntimeException}. 
Third, the server doesn't catch some bizarre scenarios, causes a Fatal Exception 
and crashes itself. In this paper, we focus on the last type of responses, as it 
is most critical and has disastrous consequences.

Depending on the implementation of the system service, the exception 
can be in the Java layer or in the native codes. A complete crash report 
consists of error messages, recorded states of the registers, stack 
traces and a memory dump. This information, especially the stack traces, 
is helpful for locating the bugs in the source codes.


\section{Testing Results and Analysis}
\label{sec:results}

Our automatic testing tool, \name, is used to test and identify vulnerable 
Android system services. We summarize the vulnerabilities identified across 
multiple Android versions, explain their security implications, and analyze 
how they survived a major open-source project like Android.

\subsection{Setups}

We tested 6 major versions of Android: 4.1 (JellyBean), 4.2 (JellyBean), 4.4 (KitKat), 
5.0 (Lollipop), 5.1 (Lollipop) and 6.0 (Marshmallow). All our experiments are conducted 
by running \nameS on official firmwares from major device manufacturers. An official 
firmware went through extensive testing by the vendors and is believed to be ready for 
a public release. Each firmware is tested in the initial state, right after it is installed. 
We didn't install any third-party app or change any configuration except for turning 
on the adb debugging option, ruling out the influence of external factors. Fig.~\ref{firmwares} 
lists the detailed information of all the firmwares we tested.

\begin{figure}[t]
\begin{center}

\small
\begin{tabular}{ccccc}
        \hline
        \textit{Version} & \textit{API} & \textit{Market} & \textit{Device} & \textit{Build \#} \\ \hline
        4.1.1 & 16 & 9.0\% & Galaxy Note 2 & JRO03C \\
        4.2.2 & 17 & 12.2\% & Galaxy S4 & JDQ39 \\
        4.4.2 & 19 & 36.1\% & Galaxy S4 & KOT49H \\
        5.0.1 & 21 & 16.9\% & Nexus 5 & LRX22C \\
        5.1.0 & 22 & 15.7\% & Nexus 5 & LMY47I \\ 
        6.0.0 & 23 & 0.7\% & Nexus 5 & MRA58K \\ \hline
\end{tabular}
\caption{\small The list of Android versions we tested using \name.}
\label{firmwares}
\end{center}
\end{figure}

An RPC method is found to be {\em vulnerable} if testing it resulted in a fatal 
exception, crashing part of, or the entire Android Runtime. Each unique crash report 
(stack traces) under an RPC interface is further referred to as an {\em individual 
vulnerability}. For each vulnerability reported here, we followed the process of: 1) 
identify it on an official ROM, 2) manually confirm that it can be reproduced, and 3) 
inspect the source codes for a root cause analysis. For vendor-specific vulnerabilities 
of which source codes are not available, such as many of the customized system services 
provided by Samsung, we only record the stack trace.

\subsection{Black-box Fuzzing Results}

We conducted a comprehensive black-box fuzzing test on 6 major versions of Android. 
Specifically, we examined more than 98 generic system services (by Google) and 72 
vendor-specific services (by Samsung), which covers more than 2400 low-level RPC methods. 
For each method, we sent either an empty transaction or a transaction filled with random 
bytes. In total, we identified 54 vulnerabilities, 39 of which are found in generic system 
services, and 15 are found in vendor-specific services. On average, each version of Android 
we tested contains 15 vulnerabilities. The latest version of Android (6.0) still contains 5 
vulnerabilities, 2 of which are new. 8 out of the 54 vulnerabilities can crash the entire 
Android Runtime (system servers), 13 can crash media servers, and 13 can cause crash of 
other system services and apps. Most of the identified bugs are due to accessing invalid 
memory addresses. We also found other causes of a crash such as StackOverflow. Fig.~\ref{exceptions} 
list all the exception types and the number of their occurrences discovered in our test.

Note that new vulnerabilities have been kept emerging on this attack surface whenever 
there is a major upgrade of Android version. We also noticed almost all of the vulnerabilities 
are found within the first few fuzzing transactions, which means that a longer fuzzing time 
did not lead to the discovery of new bugs. This suggests the inefficiency of black-box fuzzing, 
probably due to the extremely large fuzzing space. As we will show later, more vulnerabilities 
are expected if more semantically-rich fuzzing techniques are used.

\begin{figure}
\begin{center}

\small
\begin{tabular}{ clc }
        \hline
        \textit{Level} & \textit{Exception Type} & \textit{Count} \\ \hline
        \multirow{7}{*}{Java} & NullPointerException & 13 \\
        & StackOverflowError & 4 \\
        & UnsatisfiedLinkError & 2 \\ 
        & ArrayIndexOutOfBoundsException & 2 \\ 
        & OutOfResourcesException & 1 \\
        & OutOfMemoryError & 1 \\
        & StringIndexOutOfBoundsException & 1 \\
        & IOException & 1 \\ \hline
        \multirow{3}{*}{Native} & SEGV\_MAPPER & 26 \\
        & SI\_TKILL & 2 \\ 
        & SEGV\_ACCERR & 1 \\
		\hline
\end{tabular}
\caption{\small List of exceptions occurred in a comprehensive black-box fuzzing test.}
\label{exceptions}
\end{center}
\end{figure}

\subsection{Parameter-aware Fuzzing Results}

To increase the effectiveness of our testing, \nameS supports parameter-ware 
fuzzing with semi-valid transactions. Generating semi-valid transactions requires 
recording and mutating of existing valid transactions. Here, we collected more than 
one million valid transactions by running 30 popular apps in two latest Android 
versions (Android 5.1 and Android 6.0). Based on this seed dataset, we performed 
a semi-valid fuzzing test on more than 445 RPC methods of 78 system services. Note 
that we only tested the RPC methods that appeared in our seed dataset, which is a 
subset of all available RPC methods. To increase the coverage of the seed dataset, 
one can increase the duration of data collection or incorporate other data 
sources, such as the unit test cases for each system service.

We found that semi-valid fuzzing can significantly increase the scale and the 
depth of our testing. In total, we identified 89 vulnerabilities in Android 5.1 
and Android 6.0 which is 7x more than simple fuzzing. Compared to the vulnerabilities 
identified using simple fuzzing, the vulnerabilities exposed by semi-valid 
fuzzing are more interesting and have severer security implications (to be discussed later). 
Moreover, since semi-valid fuzzing is parameter-aware, we can expose different vulnerabilities 
in the same RPC method by fuzzing different parameters in the same API. For example, by 
fuzzing different variables contained in the \texttt{Intent} parameter, \nameS 
identified more than 20 vulnerabilities in a single RPC method \texttt{startActivity} 
in \texttt{ActivityManagerService}. Semi-valid testing also facilitates the process of 
identifying the corresponding bug in the source codes since we now know which input 
parameter results in the crash. Later, we will summarize the root causes of all 
identified vulnerabilities (in both simple and semi-valid fuzzing).

\subsection{Root Cause Analysis}

The direct causes of crashes are uncaught exceptions such as 
\texttt{NullPointerException} or \texttt{SEGV\_MAPPER}, but the 
fundamental cause behind them is deeper. For each crashed system 
service of which source codes are available, we looked into the source 
codes and analyzed the root causes of the vulnerabilities. Specifically, 
we are interested in why these vulnerabilities survived in a major open 
source project like Android. In summary, we found that most of the vulnerabilities 
we identified are very likely to have been overlooked by system service developers. 
A likely explanation is many system developers only considered exploitation of public 
APIs, thus directly injecting faulty transactions to the \binder\ driver creates many 
scenarios that are believed to be `unlikely' or `impossible' in their mindset. Here, 
we highlight some of the new attack vectors identified by our approach which contribute 
to most of the vulnerabilities we identified.

First, an attacker can manipulate RPC parameters even if they are not directly exposed 
via public APIs. For example, \texttt{IAudioFlinger} provides an RPC method 
\texttt{REGISTER\_CLIENT}. This method is only implicitly called in the Android 
middleware and is never exposed via public interfaces. Therefore, the developers of 
this system service may not expect an arbitrary input from this RPC method and 
didn't perform a proper check of the input parameters. In our test, sending a list 
of null parameters via the \binder\ driver can easily crash this service. This 
suggests that developers should not overlook RPC interfaces that are private or hidden.

Second, an attacker can bypass sanity checks around the public API, no matter how 
comprehensive they are. For example, the \texttt{IBluetooth} service provides a method called 
\texttt{registerAppConfiguration}. All of the parameters of this RPC method are directly 
exposed via a public API and there are multiple layers of sanity check around 
this interface. Therefore, if there is an erroneous input from the public API, the 
client will throw an exception and crash without even sending the transaction to the 
server side. However, using our approach, an attack transaction is directly injected 
to the \binder\ driver without even going through these client-side checks. This 
suggests that the server should always double-check input parameters on its own.

\begin{figure}[t]

\lstset{language=Java}

\begin{lstlisting}

android.widget.RemoteViews

private RemoteViews(Parcel parcel, BitmapCache bitmapCache) {

    int mode = parcel.readInt();

    ...

    if (mode == MODE_NORMAL) {
      ...
    } else {
      // recursively calls itself
      mL = new RemoteViews(parcel, mBitmapCache);
      // recursively calls itself
      mP = new RemoteViews(parcel, mBitmapCache);
      ...
    }

    ...    
}
\end{lstlisting}
\caption{\small The constructor of the \texttt{RemoteView} class contains a loophole 
which can cause a \texttt{StackOverflow} exception. Specifically, a bad recursion will 
occur if the input Parcel object follows a certain pattern.}
\label{remoteview}
\end{figure}

Third, an attacker can exploit the serialization process of certain data types and 
create inputs that are hazardous at the server side. For example, \texttt{RemoteView} 
is a Parcelable object that represents a group of hierarchical views. It contains a 
loophole in its de-serialization module which can cause a \texttt{StackOverflow} exception. 
As shown in Fig.~\ref{remoteview}, a bad recursion will occur if the input Parcel object 
follows a certain pattern. By directly manipulating the 
serialized bytes of the Parcel sent via the \binder\ driver, this loophole can be 
triggered and crash the server. This suggests that RPC methods with serializable 
inputs require special attention and sanity check is also essential in the 
de-serializaiton process.

\begin{figure*}
    \centering
    \hspace{5mm}
    \includegraphics[scale = 0.68]{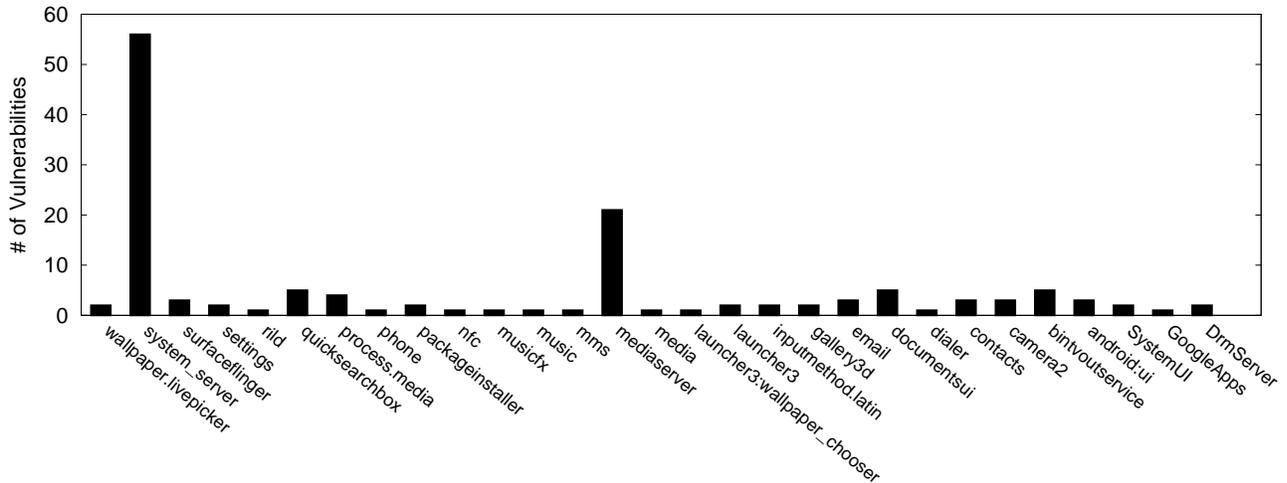}
    \caption{\small Many of the vulnerabilities we identified are found to 
    be able to crash the entire Android Runtime (system\_server), while others 
    can cause specific system services (mediaserver) or system apps (nfc, contacts, etc) to fail.}
    \label{crashes}
\end{figure*}

In summary, most of the vulnerabilities originate from unreliable 
assumptions on the transactions from the client-side. This suggests that 
there is a misconception of where the security boundary is for Android 
system services --- while many may assume the security boundary is at 
the client-side public APIs, the actual security boundary can only stand behind 
the \binder\ driver. Therefore, we advocate the importance of pure server-side 
error handling, and highlight the necessity of direct \binder-level testing 
on RPC interfaces of remote services.

\subsection{Security Implications}

Most of the vulnerabilities we discovered can be used to launch a 
Denial-Of-Service (DoS) attack. Some of them are found to be able 
to crash the entire Android Runtime, while others can cause specific 
system services or system apps to fail. Fig.~\ref{crashes} shows 
the distribution of the affected services (apps). When launching a DoS attack, 
the attacker can trigger a crash either consistently or only under certain 
conditions, for example, when a competitor's app is running. This can create the 
impression that the competitor's app is buggy and unusable. We even identified 
multiple vulnerabilities (in the de-serialization process of \texttt{Intent}) 
that can cause targeted crash of almost any system/user-level apps, without 
crashing the entire system. Specifically, an attacker can craft an Intent 
that contains a mal-formated \texttt{Bundle} object and send to the target app. 
This will cause a crash during the de-serialization process of the \texttt{Intent} 
object before the target app can conduct any sanity check. Moreover, it can be very 
challenging to identify the attacker app under these scenarios because the OS only 
knows which service/app is broken, but cannot tell who crashed it. We will discuss 
more about the attack attribution process in the next section.

Some of the vulnerabilities we discovered can cause more serious security problems. 
We found that in several RPC methods, the server-side fails to check potential 
Integer overflows. This may lead to disastrous consequences when exploited by 
an experience attacker. For example, in \texttt{IGraphicBufferProducer} an 
Integer overflow exists such that when a new \texttt{NativeHandle} is created, 
the server will malloc smaller memory than it actually requested (see Fig.~\ref{nativehandle}). 
Subsequent writes to this data struct will corrupt the heap on the 
server-side. This vulnerability has been demonstrated to be able to achieve 
privileged code execution, and insert any arbitrary code into \texttt{system\_server}~\cite{cve1528}. 
We also found a vulnerability in \texttt{IContentService} that can lead to a 
infinite bootloop, which can only be resolved by factory recovery or flushing 
a new ROM. This is also classified as High risk according to the official 
specification of Android severity levels~\cite{severity}.

\begin{figure}[t]

\lstset{language=Java}

\begin{lstlisting}

native_handle_t* native_handle_create(int numFds, int numInts)
{
    // numFds & numInts are not checked!
    native_handle_t* h = malloc( ...
        + sizeof(int)*(numFds+numInts));

    h->version = sizeof(native_handle_t);
    h->numFds = numFds;
    h->numInts = numInts;

    return h;
}

\end{lstlisting}
\caption{\small The constructor of the \texttt{native\_handle} has an Integer Overflow 
vulnerability that can cause a heap corruption on the server-side. This can lead to 
privileged code execution in \texttt{system\_server}.}
\label{nativehandle}
\end{figure}

Besides RPC methods that are not well-implemented, we also discovered RPC methods 
that are not properly protected by existing Permission models. In official ROMs of 
Samsung Galaxy 4 (Android 4.2.2 and Android 4.4.2), an attacker can reboot the device 
by directly sending a transaction to \texttt{PackageManagerService} via the \binder\ driver 
without requiring the REBOOT permission. This is critical since REBOOT 
is a sensitive permission only granted to system apps. The other service is 
\texttt{ICoverManager}, a customized service from Samsung. An attacker can 
invoke a certain RPC method of \texttt{ICoverManager} and block the entire 
screen with a pop-up blank Activity. The blank Activity cannot be revoked 
using any virtual or physical button and the only exit is restarting the device.

\subsection{Vulnerabilities: Fixed and Unfixed}

We examine how many of the vulnerabilities remain unfixed and are potentially 
zero-day when they are found. Our analysis is based on the public changes 
of the source codes across different Android versions and revisions. 
We skip the 15 vulnerabilities in vendor-specific system services 
and 7 in generic system services due to the unavailability of source codes. 
Note that not all generic system services are open source, especially when 
it is related to decryption/encryption or interactions with OEM hardware.

\begin{figure}
    \centering
    \hspace{5mm}
    \includegraphics[scale = 0.2]{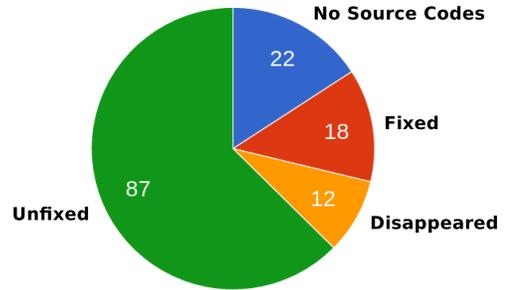}
    \caption{\small Number of the vulnerabilities that are fixed, disappeared and unfixed.}
    \label{distribution}
\end{figure}

Of the 117 analyzed vulnerabilities in Android code bases, only 18 have 
been fixed by adding additional sanity checks of input parameters. Another 12 
vulnerabilities `disappeared' during several major Android version upgrades 
either because 1) the corresponding source codes (or API) have been deleted; 
or 2) new updates in other parts of the source codes accidentally bypass the 
vulnerable source codes. For example, some crashes are caused by a recursive 
call in the \texttt{RemoteView} class (see Fig.~\ref{remoteview}). Similar 
crashes disappeared after Android 5.0. We looked in the source codes and found 
this is not because the bug has been fixed, but because in new versions of Android 
a faulty transaction will create an additional Exception before it reaches the vulnerable 
codes. The additional Exception is properly caught and accidentally avoids the fatal 
crash caused by the real vulnerability. We do not consider this as a `fix' since an 
attacker can still recreate the crash by manually crafting a transaction which bypass 
the new code updates. As of this writing, there are still 87 vulnerabilities left unfixed. 
Fig.~\ref{distribution} illustrates the proportion of vulnerabilities that are fixed, 
disappeared and unfixed. We have already submitted all unfixed vulnerabilities to AOSP 
by the time of this submission.

\section{Defenses}

As our testing results demonstrated, new vulnerabilities emerge whenever 
there is a major upgrade of the Android code base. This is because, considering the 
code size of Android, it is almost impossible to prevent the developers from 
writing buggy codes. Therefore, the only solution is to eliminate potential 
bugs as early as possible in the development cycle. To this end, we discuss 
potential defense mechanisms and summarize them into two categories: (1) precautionary 
testing, exposing vulnerabilities before releasing the new ROM; (2) runtime 
defense, defending against potential bugs after the ROM has been deployed. Next, 
we will go through several potential defense mechanisms in each category and discuss 
whether it is applicable or practical for our problem. Then, we demonstrate how 
to enhance the visibility of ongoing attacks by enabling runtime diagnostic of 
\binder\ transactions.

\subsection{Precautionary Testing}

Before releasing a new ROM, developers can conduct precautionary testing. 
The defense can be done early, in the development phase of each system 
service, or later, after the entire ROM gets built.

Android has already adopted a static code analysis tool, \texttt{lint}, to check 
potential bugs and optimizations for correctness, security, performance, 
usability, accessibility and internationalization~\cite{lint}. Specifically, 
\texttt{lint} provides a feature that supports inspection with annotations. 
This allows the developer to add metadata tags to variables, parameters and 
return values. For example, the developer can mark an input parameter as 
\texttt{@NonNull}, indicating that it cannot be \texttt{Null}, or mark it 
as \texttt{@IntRange(from=0,to=255)}, enforcing that it can only be within 
a given range. Then, \texttt{lint} automatically analyzes the source codes 
and prompts potential violations. This can be extended to support inspections 
of RPC interfaces, allowing developers to explicitly declare the constraints 
for each RPC input parameter. This way, many potential bugs can be eliminated 
during the development phase. This defense is practical and comprehensive 
but requires system developers to specify the metadata tags for each RPC interface.

We can also conduct precautionary testing during runtime after the ROM has 
been built. Our system, \name, itself is effective in identifying vulnerabilities 
and can be used as an automatic testing tool. By fuzzing various system services 
with different policies, a large number of vulnerabilities can be eliminated 
before reaching the end-users. Actually, many severe vulnerabilities~\cite{peles2015one,cve1528,cve1474} 
could have been avoided if \nameS had been deployed. Note that the effectiveness 
of \nameS depends on the quality and coverage of the seed transactions. Besides 
collecting execution traces of a large number of apps, another potential way of 
generating a comprehensive seed dataset is to incorporate the functional unit 
tests of each system service.

\subsection{Runtime Defense}

It will be helpful if Android can provide some real-time defense against 
potential vulnerabilities even after the ROM has been deployed on end-users' devices. 
Here, we focus on specific defenses on the \binder\ layer, excluding 
generic defenses such as Address Space Layout Randomization (ASLR), SELinux, 
etc. They have been extensively discussed in other literature 
\cite{Lee:2014:ZMF:2650286.2650762,shabtai2009securing,smalley2013security}
and are not specific to our scenario. There are two potential defenses one 
can provide on the \binder\ surface during runtime: (i) intrusion detection/prevention, 
identifying and rejecting transactions that are malicious, and (ii) intrusion 
diagnostics, making an attack visible after the transaction has already 
caused some damage.

To provide runtime intrusion prevention, one needs to perform some type 
of abnormality detection on incoming transactions. This works by examining the 
input parameters of valid/invalid RPC invocations and characterizing the 
rules or boundaries. However, in our case, it is not practical for the 
following reasons. First, \binder\ transactions occur at a very high frequency 
but a mobile device is itself constrained in energy and computation power. Second, 
parameters in \binder\ transactions are very diverse, codependent, and evolving 
dynamically during runtime, and hence clear boundaries or rules may not exist. 
Third, end-users are not likely to endure even the smallest false-positive rate. 
One can, of course, build a very conservative blacklist-based system and hard-coding 
rules of each potential vulnerability in the database. However, this seems unnecessary, 
especially when Android nowadays supports directly pushing security updates (patches) 
to devices of end-users.

An alternative solution is to diagnose, instead of prevent. It would be helpful 
if we can provide more visibility of how malicious transactions actually 
undermine a device. Even though this cannot stop the single device from being 
attacked, we can still utilize the collected statistics to develop in-time 
security patches, benefiting the vast majority of end-users. According to our 
experience in bug-hunting process (in Section 6), improvements on the following 
two aspects can effectively increase the visibility of \binder\ transactions.

\paragraph{Construct IPC Graph.}
In the case of an attack on the \binder\ interface, the victim is the recipient of 
the transaction while the attacker is the sender. Under most circumstances, there 
will only be visible consequences (i.e., crashes) on the server side while the 
attacker stays in the mist. This makes it difficult to conduct attack attribution, 
especially when an attack is mounted via a chain of transactions across multiple 
processes. To enhance the visibility of attacks in this process, we can instrument 
each system service to maintain the senders of recent transactions. Each transaction 
represents an edge in the IPC graph, linking apps and system services. Any user-level 
app that is linkable to the victim system service is a potential initiator of the 
attack. Similar techniques have been proven to be effective in improving the visibility 
of remote systems~\cite{dietz2011quire}.

\paragraph{Maintain Transaction Schema.}
Even if we know which transaction causes a crash, it is often challenging to 
identify the corresponding vulnerability, especially when the bug is non-trivial. 
The most frequent obstacle is lack of visibility of the transaction schema. 
A transaction contains only raw bytes and may be unmarshalled in any arbitrary way. 
It is extremely tedious to anatomize a transaction and verify that it actually contains an 
invalid parameter causing a specific vulnerability. We propose to maintain transaction 
schema at runtime when the system service parses a transaction (similar to Fig.~\ref{intent}). 
If the server experiences some exception, the recorded schema will be attached with the crash 
report to provide more informative feedback.


\section{Discussion}

Our work has lasting values beyond the vulnerabilities we presented in this 
paper. First, we comprehensively assessed a risky attack surface that 
has long been overlooked by the system developers of Android. As our 
experimental results demonstrated, new vulnerabilities are still emerging on 
this attack surface and \nameS can help eliminate potential vulnerabilities in future 
releases of Android.

Second, the lessons learned can transcend to other platforms facing similar 
issues, such as vehicular systems (CAN buses and ECUs), wearable devices, 
etc. We highlight that, although many systems adopt a client--server model 
in the design of their internal system components, they rarely follow the 
security standards of a real client--server model as in a networked environment. 
In many scenarios, a component may fall into the wrong hands and create serious 
security threats.

Third, our parameter-aware fuzzing is generic and not limited to system services. 
In fact, it also works for services exported by user-level apps. For example, 
apps like Facebook also host service in its own process space and export it 
to other apps. By performing fuzzing on this interface, more app-level vulnerabilities 
are expected to be unearthed. We didn't discuss it here mainly because source codes 
of app-level services are mostly unavailable. Therefore, it is difficult for us to 
analyze the root causes and security implications of the identified vulnerabilities.

For two reasons, we test each firmware in its initial state: right after it has been flushed, 
without installing third-party apps, inserting SIM cards, or connecting to WiFi. First, 
we want to exclude vulnerabilities caused by external factors that may not be reproducible. 
Second, we want to rule out the possibility that our fuzzing tests would negatively affect 
cellular providers or Internet services, for the sake of responsible research. All of the 
vulnerabilities we identified have been reported to AOSP. Part of the vulnerabilities have 
been accepted and will be patched in future versions of Android, while the rest are still 
under review. In this paper, we have only revealed details about the vulnerabilities that 
have been confirmed so far.


\section{Conclusion}

In this paper, we have conducted a field study accessing the robustness of 
Android system services. Specifically, we have designed and implemented \name, 
an automatic testing framework that can help expose vulnerable system services 
by fuzzing the \binder\ interface. \nameS supports parameter-aware 
fuzzing and identified more than 100 vulnerabilities in 6 major 
versions of Android. We summarized these vulnerabilities, explained 
their security implications and analyzed their root causes. Based on 
our observation, we highlighted the deficiency of testing only on 
client-side public APIs and advocated testing and protection at 
the \binder\ interface --- the actual security boundary. Several 
potential defenses as well as their practicality have been discussed to 
help eliminate vulnerabilities as early as possible in the development cycle.


{\footnotesize \bibliographystyle{acm}
\bibliography{sigproc}}

\end{document}